\documentclass[doublespacing]{elsart}

\usepackage{mathenv} %needed to define 'd' column type
\colset{tabular}              %needed to define 'd' column type
\usepackage{dcolumn} %needed to define 'd' column type
\newcolumntype{d}[1]{D{.}{.}{#1}}%define 'd' column type (align at decimal point)

\usepackage{graphicx}% Include figure filES

\begin{document}
\begin{frontmatter}
\title{A new challenge for time-dependent density-functional theory} \author{Meta van Faassen and Kieron Burke}
\address{Department of Chemistry and Chemical Biology, Rutgers University, Piscataway, New Jersey 08854-8087}
\begin{abstract}
  Time-dependent density functional theory is thought to work well for
  the test cases of He and Be atoms.  We perform a quantum defect
  analysis of the $s$, $p$, and $d$ Rydberg states of Be with accurate
  ground state Kohn-Sham potentials. The $s$ and $p$-quantum defects are well
  described by the ALDA, but fails
  badly for the $d$-quantum defect. The same failure is observed in
  case of He. This provides a new challenge for functional development
  in time-dependent density functional theory.
\end{abstract}
\end{frontmatter}
\section{Introduction}
Time-dependent density functional theory (TDDFT)~\cite{RG84} has become a
popular method for calculating excitation energies of atoms,
molecules, clusters, and solids~\cite{FR05,BWG05}. In many common
situations, it yields transition frequencies with errors below 0.2 eV,
and properties of excited states are obtained with an accuracy
comparable to that of ground-state DFT.

As any theory is increasingly applied, the limitations of its
approximations are discovered~\cite{BWG05}.  An early limitation for
TDDFT was due to the poor asymptotic behavior of potentials in
approximations to ground state DFT. This problem was overcome by the
development of asymptotically corrected
potentials~\cite{LB94,CCS98,LRC95,WAY03}. Later, it was found that
adding an asymptotic tail to the LDA potential, without any
modification in the inner region, gives very accurate
results~\cite{WB05}. Another limitation was that within the adiabatic
approximation, double excitations are missed by TDDFT. It was shown by
Maitra {\em et al.} that this problem can be overcome with a frequency
dependent exchange-correlation (xc) kernel~\cite{MZCB04,ZB04}. A still
open question is how to obtain charge-transfer excitations within
TDDFT. An empirical approximation of the xc-kernel has been designed
as a step to overcome this problem~\cite{GB04}.

In this communication, we report a new and surprising failure of
present TDDFT calculations.  While previously hinted at in the
literature, our recently-developed quantum defect analysis (QDA) for
atoms~\cite{FB06} makes it very clear that this is a qualitative
failure.  Essentially, as the angular-momentum change grows, TDDFT
with the adiabatic local density approximation (ALDA), becomes worse
and worse. For $s\to d$ transitions, including
TDDFT xc-contributions is much worse than
ignoring them altogether!
\section{Theory}
For closed shell atoms and for any spherical one-electron potential
that decays as $-1/r$ at large distances, the bound-state transitions
form a Rydberg series with frequencies:
\begin{equation}
\omega_{nl}=I-\frac{1}{2(n-\mu_{nl})^2}
\label{eq:qd}
\end{equation}
where $I$ is the ionization potential, and $\mu_{nl}$ is called the
quantum defect~\cite{F98}.  We use atomic units ($e^2=\hbar=m_e=1$) throughout.
For real atoms, quantum
defects depend only weakly on the principal quantum number $n$ for
large $n$ and converge to a finite value in the limit
$n\rightarrow\infty$. 

The quantum defect is a smooth function of energy, and is well
approximated a polynomial of some low order $p$:
\begin{equation}\label{eq:fit}
\mu^{(p)}(E) =\sum_{i=0}^p \mu_i E^i,\quad E=\omega-I.
\end{equation}
Thus we calculate transition frequencies, extract the quantum defect
via Eq.~\ref{eq:qd} and plot them as a function of $E$.  When we fit
our quantum defect values to Eq.~\ref{eq:fit}, we optimize the fit
over the entire range of excitation energies, not just around $E=0$
(the $\mu_i$ are simply related to the $a$, $b$, and $c$ coefficients
in Refs.~\cite{ARU98} and ~\cite{AQR99}).

For the TDDFT part, we solve the linear response equations as derived
by Casida~\cite{C95}. Such a TDDFT calculation consists of two parts.
First the ground state KS orbitals and KS orbital energies are
determined with DFT. For He and Be, nearly exact potentials have been
developed by Umrigar {\em et al.}  (He~\cite{UG94} and
Be~\cite{UG93}), and we use these potentials for the ground state. In
this way, we eliminate any error due to the underlying ground state KS
potential.  In the second part of the calculation (the TDDFT part), we
do perturbation theory on the ground state results. This means solving
Casida's equations, and we use the ALDA xc-kernel in all cases.  We
also include results in the Hartree approximation, equivalent to
setting the xc-kernel to zero. In this approximation triplet values
correspond to the bare KS orbital energy differences, as the Hartree
interaction involves no spin flips.
\section{Computational details}
For the calculation of KS orbital energies and TDDFT excitation
energies, we used the Amsterdam Density Functional program package ADF
(http://www.scm.com). The numerical integration accuracy was set to 12
significant digits, and the convergence criterion in the
self-consistent procedure for the solution of the KS equations was set
to $10^{-12}$. We reached convergence with a large even tempered basis
set with primitive basis functions including $f$ functions and a fit
set including $g$ functions.

In order to map the accurate potentials onto the integration grid of
ADF we performed a cubic spline interpolation between the available
data points. We confirmed the quality of our ground state KS results
by comparing the ADF orbital energies with orbital energies obtained
with an OEP (optimized potential model)
program~\cite{Engel,ED99,TS76}. This program is basis set independent,
works with a radial grid, and both the energies and the potentials are
optimized in a self-consistent way. For this code, the accurate
potentials do not need to be interpolated. We obtained ground state
orbitals energies with ADF that were identical to the ones obtained
with the OEP code at least up to 5 digits (in a.u.).

In order to obtain the coefficients in Eq.~\ref{eq:fit} for He, we
included states up to $n=6$ in case of the ALDA and $f_{\rm xc}=0$
calculations, up to $n=10$ in case of the accurate wave function
reference data, and up to $n=7$ for the KS values.  For Be we included
states up to $n=6$ in case of ALDA and $f_{\rm xc}=0$ calculations, up
to $n=8$ in case of the experimental reference data, and up to $n=9$
for the KS values.
\section{Results}
Van Gisbergen {\em et al.} first calculated transitions from the
accurate Be KS potential~\cite{GKSG98} using the ALDA kernel and
reported tables with transitions. If the same data is cast into a
quantum defect and plotted as function of energy, systematic errors
are revealed and more insight in the results is obtained. We also did
calculations on higher $s\to d$ transitions that have not been studied
previously.

The continuous lines in all graphs we show correspond to the
fitted quantum defect values. The coefficients used to obtain the
continuous quantum defect, using Eq.~\ref{eq:fit}, are reported in
Ref.~\cite{FB06} for the $s$- and $p$-cases. 
\begin{figure}
\includegraphics{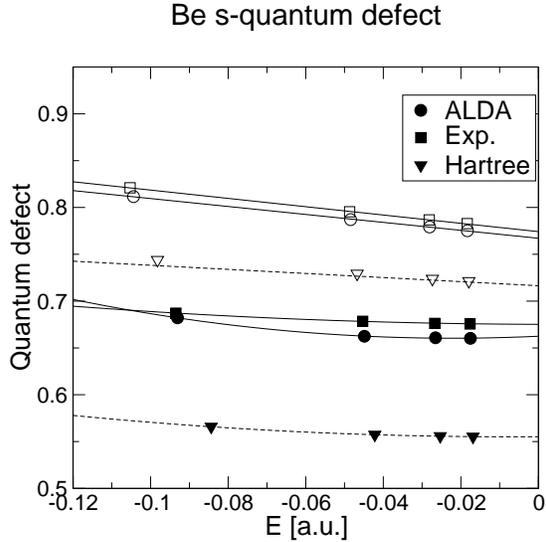}
\caption{\label{fig:besplot}Hartree and ALDA singlet and triplet $s$-quantum 
  defects of Be, compared with experimental values~\cite{KM97}.  
The closed symbols correspond to singlet values, the open symbols correspond to triplet values.}
\end{figure}
We show the $s$-quantum defect of Be in Figure~\ref{fig:besplot}.  The
dashed line with open triangles shows the scattering from the ground
state KS potential, which lies neatly between the experimental singlet
(closed squares) and triplet (open squares) values. We note how large
the quantum defects are. We see from Eq.~\ref{eq:qd} that a quantum
defect of 1 means (for single-particle scattering) that the deviation
from $-1/r$ is so big that the energy of a particular transition to a
state with quantum number $n$ corresponds exactly to the energy of
that transition in a hydrogen atom to a final state with quantum
number $n'=n-1$.  Inclusion of only Hartree effects shifts this line
down to the closed triangles for the singlet, but leave the triplets
unchanged. Finally, the ALDA xc-kernel shifts both of these Hartree
results upward to the circles. The ALDA accounts for more that 80\% of
the error of the Hartree approximation.

\begin{figure}
  \includegraphics{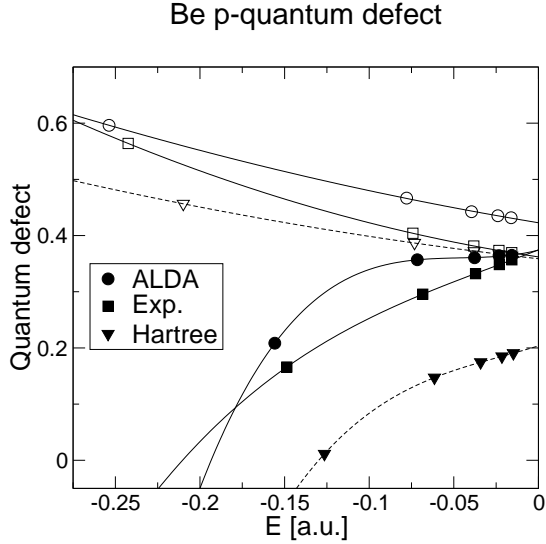}
\caption{\label{fig:bepplot}Hartree and ALDA singlet and triplet $p$-quantum 
  defects of Be, compared with experimental values~\cite{KM97}.  
The closed symbols correspond to singlet values, the open symbols correspond to triplet values.}
\end{figure}
We next show the the $p$-quantum defect in Figure~\ref{fig:bepplot}.
In this case we are seeing a lot more structure than in case of the
$s$-quantum defect. We see that the ALDA slightly overestimates the
curvature of the quantum defect correction in this case. This produces
the largest errors for the most negative energies (i.e. the lowest
transitions). When $E$ goes to zero all quantum defect values get
close together except for the singlet Hartree values which
underestimate the experimental results.  The triplet Hartree results
are a good approximation to the experiment. Here ALDA clearly
overestimates the correction to $\mu(0)$. This is not a large
problem for lower transitions, but causes the $E\to0$ results to be
worse than bare Hartree. In fact, the KS values lie in-between the
experimental singlet and triplet values until $n=6$. It can be seen
from the graph that they do seem to get below the experimental singlet
and triplet results for $n>6$. Indeed, for $n=7$, both the singlet and
triplet experimental quantum defects (0.3616 and 0.3676 respectively)
lie above the KS value of 0.3550.  So here we have an example where
the quantum defect of the bare KS orbital energy differences does not
lie between the experimental singlet and triplet values.

\begin{figure}
\includegraphics{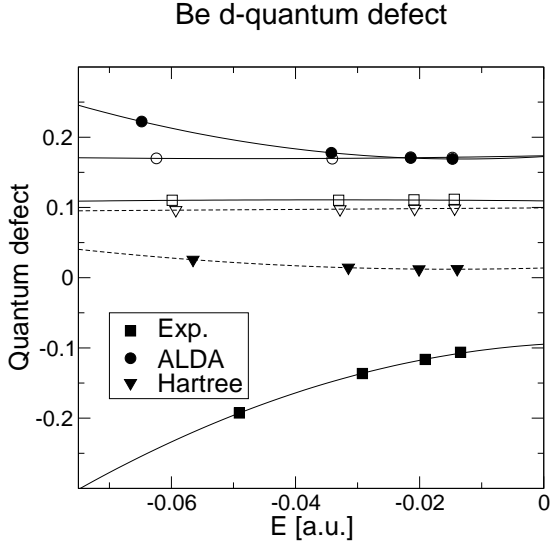}
\caption{\label{fig:bedplot}Hartree and ALDA singlet and triplet $d$-quantum 
defects of Be, compared with experimental values~\cite{KM97}.  
The closed symbols correspond to singlet values, the open symbols correspond to triplet values.}
\end{figure}
\begin{figure}
\includegraphics{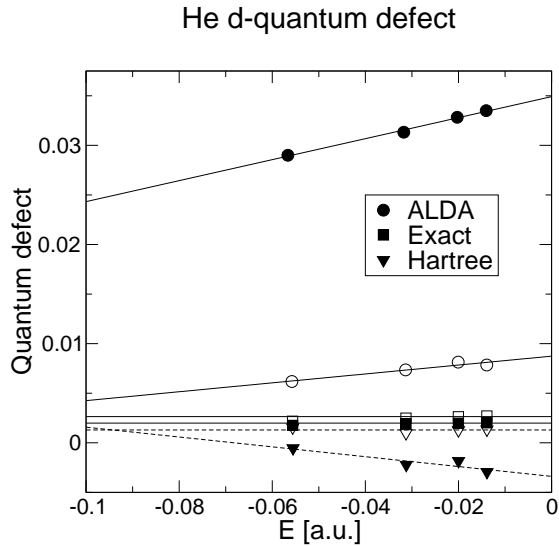}
\caption{\label{fig:hedplot}Hartree and ALDA singlet and triplet $d$-quantum 
defects of He, compared with accurate wave function results~\cite{D96} (labeled ``Exact''). 
The closed symbols correspond to singlet values, the open symbols correspond to triplet values.}
\end{figure}
We show the $d$-quantum defect of Be in Figure~\ref{fig:bedplot} and
that of He in Figure~\ref{fig:hedplot}. For Be we first note that the
KS curve lies, as before, between the singlet and triplet. But in this
case it is much closer, and parallel to, the triplet curve. In this
case the Hartree approximation does rather well, by shifting the
singlet curve about halfway towards the experiment, while not moving
the triplet. But the ALDA curves are a disaster. For the singlet, the
ALDA correction is in the wrong direction! For the triplet the
correction is in the right direction, but it is too large by a factor
of about four. We find similar results in case of He
(Figure~\ref{fig:hedplot}). First of all, the KS curve lies below the
singlet and triplet curves for all $n$ values studied. Again it is
close to the triplet curve, the singlet Hartree curve underestimates
the reference curve. The ALDA correctly shifts both singlet and
triplet curves upward, but the shift is too large for the singlet,
again leading to a dramatic overestimation.

\begin{table}
\caption{\label{tab:hebeqd} We give the coefficients of the expansion of the 
singlet and triplet $d$-quantum defect of He and Be. 
The ionization energy of He is 0.9037 a.u., and that for Be is 0.3426 in all cases.}
{\begin{tabular}{cc|d{4}d{4}|d{4}d{4}|d{4}d{4}} \hline
&         &  
\multicolumn{2}{c|}{Reference\footnotemark[1]} &
                \multicolumn{2}{c|}{$f_{\rm xc}=0$}  & 
                \multicolumn{2}{c}{ALDA} \\
Atom & & \multicolumn{1}{c}{Singlet} &  \multicolumn{1}{c|}{Triplet} &
\multicolumn{1}{c}{Singlet} & \multicolumn{1}{c|}{Triplet} & 
\multicolumn{1}{c}{Singlet} & \multicolumn{1}{c}{Triplet} \\\hline
He  & $\mu_0$  & 0.0020   & 0.0026  & -0.0034 & 0.0013    & 0.0349   & 0.0087 \\
    & $\mu_1$  &          &         & -0.0496 &           & 0.1059   & 0.0449\\ 
    & Max. AE  & 0.0002   & 0.0005  & 0.0006 & 0.0003     & 0.0002 & 0.0003\\\hline 
Be  & $\mu_0$  & - 0.0947 &  0.1094 & 0.0138 & 0.0994    &  0.1725  & 0.1736\\
    & $\mu_1$  &   0.5636 & -0.0810  & 0.2353 &  0.0554  &  0.5291  & 0.1699 \\ 
    & $\mu_2$  & -29.2439 & -1.1540 & 7.8553 &          & 20.0356  & 1.7686 \\
    & Max. AE  &   0.001  &  0.001  & 0.0003 & 0.0002  & 0.0000   & 0.0008 \\\hline
\end{tabular}}
\footnotemark[1] {Accurate wave function results for He from Ref.~\cite{D96} 
and experimental data for Be from Ref.~\cite{KM97}}.
\end{table}
We show the coefficients of the $d$-quantum defect corresponding to
the fit of Eq.~\ref{eq:fit} in Table~\ref{tab:hebeqd}. The columns
labeled KS and Hartree are both referring to the Hartree
approximation, we label the triplet data by KS to emphasize that these
values correspond to the bare KS orbital energy differences.  Apart
from the coefficients we also give the ``Max.  AE'' in this tables, by
which we mean the maximum absolute error between the fitted values and
the true quantum defect values.  In all cases we stopped adding
coefficients to the fit expansion until this error was smaller than
0.001, or when adding more coefficients no longer reduced the error.
We see that three coefficients are enough in all cases to completely
describe the quantum defect.  The ALDA estimate for the $E=0$ quantum
defect ($\mu_{0}$) is close to the exact data for $l=0,1$, but not for
$l=2$ and we also see that the $E=0$ quantum defect is not always
between the exact singlet and triplet values.

Thus in case of the $d$-quantum defect, including the ALDA kernel
within TDDFT worsens the results instead of improving it. In the mid
1980s a similar failure of the LDA was observed in the case of
transfer energies~\cite{GJ85,GJb85}. It was found that $sp$ transfer
energies in first row atoms and $sd$ transfer energies in $3d$ atoms
are underestimated by the local spin-density approximation (LSD). The
source of the error was linked to the fact that the LSD is unable to
properly take into account the nodal structure of the orbitals. The
failure of the ALDA that we observe here may be related to the LSD
problem, but the methodology here is very different.  In TDDFT, we
extract transitions by using linear response, while in order to obtain
transfer energies, one subtracts total energy differences.

\begin{figure}
\includegraphics{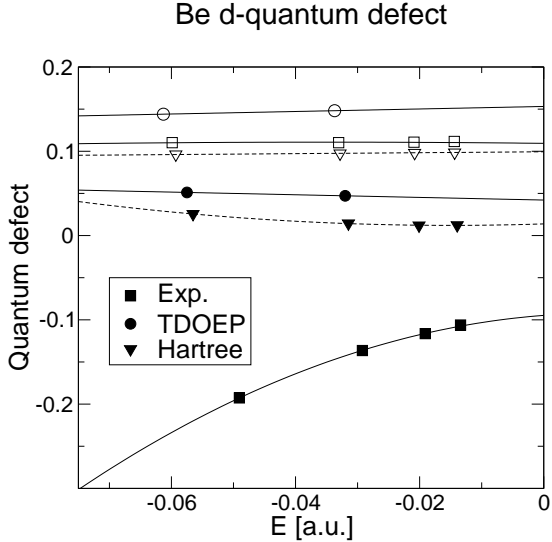}
\caption{\label{fig:bedTDOEP} Hartree and TDOEP singlet and triplet $d$-quantum 
defects of Be, compared with experimental values~\cite{KM97}. 
The closed symbols correspond to singlet values, the open symbols correspond to triplet values.}
\end{figure}
To test the effect of including orbital dependence in the kernel we
extracted quantum defect values from the exchange-only OEP
calculations of Petersilka {\em et al.}~\cite{PGB00}. They did not do
full time-dependent OEP, instead they use the approximate PGG
kernel~\cite{PGG96}.  We show results in Figure~\ref{fig:bedTDOEP}.
The errors are similar to those of ALDA, but considerably
quantitatively smaller. A full OEP calculation might do better still.

This work was supported by NSF Grant No. CHE-0355405.
%
%\bibliography{myrefers}% Produces the bibliography via BibTeX

\end{document}